\documentclass[twocolumn,showpacs,aps,prl,]{revtex4-1}
\usepackage{verbatim} 
\usepackage{graphicx}
\usepackage{dcolumn}
\usepackage{bm}
\usepackage{array}
\usepackage{booktabs}
\usepackage{amsmath}
\usepackage{color}
\usepackage{times}
\usepackage{epstopdf}
\usepackage{lineno}
\begin{document}

\title{
Observation of di-structures in $e^+e^-\rightarrow{J}/\psi{\rm X}$ at center-of-mass energies around 3.773 GeV
}
\author{
\begin{small}
\begin{center}
M.~Ablikim$^{1}$, M.~N.~Achasov$^{10,e}$, P.~Adlarson$^{63}$, S.
~Ahmed$^{15}$, M.~Albrecht$^{4}$, A.~Amoroso$^{62A,62C}$, Q.~An$^{59,47}$,
~Anita$^{21}$, Y.~Bai$^{46}$, O.~Bakina$^{28}$
% ...
\\
\vspace{0.2cm}
(BESIII Collaboration)\\
\vspace{0.2cm} {\it
$^{1}$ Institute of High Energy Physics, Beijing 100049, People's Republic
of China\\
% ...
$^{4}$ Bochum Ruhr-University, D-44780 Bochum, Germany\\
% ...
$^{10}$ G.I. Budker Institute of Nuclear Physics SB RAS (BINP), Novosibirsk
630090, Russia\\
% ...
$^{15}$ Helmholtz Institute Mainz, Johann-Joachim-Becher-Weg 45, D-55099
Mainz, Germany\\
% ...
$^{21}$ Indian Institute of Technology Madras, Chennai 600036, India\\
% ...
$^{28}$ Joint Institute for Nuclear Research, 141980 Dubna, Moscow region,
Russia\\
% ...
$^{46}$ Southeast University, Nanjing 211100, People's Republic of China\\
$^{47}$ State Key Laboratory of Particle Detection and Electronics, Beijing
100049, Hefei 230026, People's Republic of China\\
% ...
$^{59}$ University of Science and Technology of China, Hefei 230026,
People's Republic of China\\
% ...
$^{62}$ (A)University of Turin, I-10125, Turin, Italy; (B)University of
Eastern Piedmont, I-15121, Alessandria, Italy; (C)INFN, I-10125, Turin,
Italy\\
$^{63}$ Uppsala University, Box 516, SE-75120 Uppsala, Sweden\\
% ...
$^{e}$ Also at the Novosibirsk State University, Novosibirsk, 630090,
Russia\\
% ...
      }
    \end{center}
    \vspace{0.4cm}
  \end{small}
}
\noaffiliation
\begin{abstract}
We report a measurement of the observed
cross sections of the inclusive $J/\psi$ production in 
$e^+e^-\rightarrow {J}/\psi{\rm X}$
based on 3.21~fb$^{-1}$ of data accumulated
at energies from 3.645 to 3.891~GeV with the BESIII detector operated
at the BEPCII collider.
The energy-dependent lineshape obtained from the measured cross sections
cannot be well described by two Breit-Wigner (BW) amplitudes
of the expected decays $\psi(3686)\rightarrow {J}/\psi{\rm X}$ and $\psi(3770)\rightarrow {J}/\psi{\rm X}$.
Instead it can be better described with three BW amplitudes
of the decays $\psi(3686)\rightarrow {J}/\psi{\rm X}$,
$R(3760)\rightarrow {J}/\psi{\rm X}$ and $R(3790)\rightarrow {J}/\psi{\rm X}$
with two distinct structures referred to as $R(3760)$ and $R(3790)$.
Under this assumption, we extracted their masses, total widths, and the product of the leptonic width
and decay branching fractions to be
$M_{R(3760)}= {3761.7\pm 2.2 \pm 1.2}$ MeV/$c^2$,
$\Gamma^{\rm tot}_{R(3760)}= {6.7\pm 11.1 \pm 1.1}$ MeV,
$\Gamma^{ee}_{R(3760)}\mathcal B[R(3760)\rightarrow {J}/\psi {\rm X}]=(4.0\pm 4.3\pm 1.2)$ eV,
$M_{R(3790)} = {3784.7\pm 5.7 \pm 1.6}$ MeV/$c^2$, 
$\Gamma^{\rm tot}_{R(3790)} = {31.6 \pm 11.9 \pm 3.2}$ MeV,
$\Gamma^{ee}_{R(3790)}\mathcal B[R(3790)\rightarrow {J}/\psi {\rm X}]=(18.1\pm 10.3\pm 4.7)$ eV,
where the first uncertainties are statistical and second systematic.
\end{abstract}


\maketitle

The mesons with mass  above the threshold of open charm (OC) pairs 
had been considered for more than 25 years to decay entirely to
OC final states via the strong interaction.  Only a few experimental 
studies of non-OC (NOC) decays of these mesons had been carried out 
before the summer of
2002~\cite{BES_intial_psi3770_physocs_program,BESIII_eeToMuMu_arXiv2007p12872v1_hep_ex_25Jul2020}.
In July 2003, the BES Collaboration claimed for the first time that they had observed $7\pm3$ events of
the NOC final state of ${J}/\psi\pi^+\pi^-$~\cite{arXiv_hep_ex_0307028v1}
in the $e^+e^-$ collision data taken with the BES-II detector operated at the BEPC collider
at center-of-mass energies nearby 3.773 GeV. This observation started world-wide a new era
with the aim to study rigorously NOC decays of the mesons lying above open-charm thresholds.
After more than two years of 
intensive discussion in the particle physics community about whether
${J}/\psi\pi^+\pi^-$ is really a decay product of the mesons lying
above the lowest open-charm threshold (3.73 GeV), it has been accepted
that this golden final state is a product of the $\psi(3770)$
NOC decays.
However, it has not been excluded that this golden final state
may be a decay product of some other possible structures~\cite{bes2_prl_2structures}
which was speculated to exist in this energy region.
The discovery of the first NOC final state of $J/\psi\pi^+\pi^-$ from the meson(s) decays
overturns the conventional knowledge that
almost $100\%$ of the mesons decay into OC final states through the strong
interaction. It stimulated a strong interest in studying NOC
decays of other mesons lying above the OC thresholds
and it inspired more experimental efforts to study NOC
decays of the mesons. In particular, the study of the ${J}/\psi\pi^+\pi^-$ final state
or a similar final state such as ${\rm M_{c\bar c}X_{LH}}$
(${\rm M}_{c\bar c}$ is a hidden charm meson such as $J/\psi$, $\psi(3686)$, $\chi_{cJ}{(J=0,1,2)}$
and $h_c$ ..., while ${\rm X_{LH}}$ refers to any allowed light hadron(s))
lead to the discovery of several new states
~\cite{X3872_PhysRevLett91_262001_Y2003, X4260_PRL95_142001_Y2005},
such as the historically labeled $X$, $Y$, and $Z$ states.

According to the potential model~\cite{Eithtin_chmonuim_prd1978}, the
$\psi(3770)$ resonance is the only $c\bar c$ state which can be directly
produced in $e^+e^-$ annihilation in the energy region between 3.73 and 3.87~GeV.
The $\psi(3770)$ resonance is
expected to decay to $D\bar D$ meson pairs
with a branching fraction of more than $99\%$,
and to decay to $e^+e^-$ and $\gamma\chi_{cJ}$
($J=0,1,2$) with a total branching fraction of less than $1\%$~\cite{Eithtin_chmonuim_prd1978}.
However, the BES Collaboration found large fractions of the
$\psi(3770)$ decaying to non-${D\bar D}$ using different data samples
which are
$(14.5\pm 1.7 \pm 5.8)\%$~\cite{PhysLettsB641_145_2006},
$(16.4\pm 7.3\pm 4.2)\%$~\cite{PhysRevLetts97_121801_2006},
$(13.4\pm 5.0\pm 3.6)\%$~\cite{PhysRevD76_122002_2007}, and
$(15.1\pm 5.6\pm 1.8)\%$~\cite{PhysLttesB659_74_2008}.
These large branching fractions for $\psi(3770)$ decaying to non-${D\bar D}$
indicate that the $\psi(3770)$ may be not a pure $c\bar c$ state or
due to the presence of some unknown structure(s) lying at energies
nearby $\psi(3770)$~\cite{RongG_CPC_34_778_Y2010}.
To search for the new structure(s), 
as suggested in Ref.~\cite{RongG_CPC_34_778_Y2010},
we studied the processes $e^+e^-\rightarrow {J}/\psi{\rm X}$ 
in the energy region between 3.645 and 3.891~GeV, since an analysis of
cross sections for $e^+e^-\rightarrow f$, whereby $f$ refers to any final state,
at energies above the $f$ threshold can directly reveal new
states~\cite{RongG_CPC_34_778_Y2010,Ablikim:2019hff}.

In this Letter, we report a measurement of the observed cross sections for $e^+e^-\rightarrow {J}/\psi{\rm X}$
based on 3.21~fb$^{-1}$ of data taken at 69 center-of-mass (c.m.) energies ranging from 3.645 to 3.891~GeV.
These data were accumulated with the BESIII~\cite{bes3} detector at the BEPCII~\cite{bes3} collider,
which comprise to integrated luminosity of 72~${\rm pb}^{-1}$ 
of cross-section scan data~\cite{Lum_72inspb_Psi3770ScanData},
44.5~${\rm pb}^{-1}$ taken at 3.650~GeV, 162.8~${\rm pb}^{-1}$ taken at 3.6861~GeV~\cite{Lum_162p8inspb},
2.93~${\rm fb}^{-1}$ taken at 3.773 GeV~\cite{Lum_2931inspb_Psi3770Data_PLB753_629},
and 50.5~${\rm pb}^{-1}$ taken at 3.808~GeV.

The BESIII detector and its response are described elsewhere~\cite{bes3_simulation}.
Here, we discuss only those aspects that are specifically related to this study.
The production of the $\psi(3686)$ and $\psi(3770)$ resonances are simulated with the
Monte Carlo (MC) event generator {\sc kkmc}~\cite{kkmc}.
The decays of these resonances to
${J}/\psi \pi\pi$, ${J}/\psi \eta$, ${J}/\psi \pi^0$, and $\gamma\chi_{cJ}$ ($J=0,1,2$)
are generated with {\sc EvtGen}~\cite{besevtgen} according to the relative branching fractions
of these final states~\cite{pdg2018}.
To study possible backgrounds,
MC samples of inclusive $\psi(3686)$ and $\psi(3770)$ decays,
$e^+e^-\to(\gamma)J/\psi$, $e^+e^-\to(\gamma)\psi(3686)$,
$e^+e^-\to q\bar q$ ($q=u,d,s$), and other final states which may be misidentified as
${J}/\psi{\rm X}$ are also generated,
where $\gamma$ in parentheses denotes the inclusion of photons from
the Initial State Radiation (ISR).

The observed cross section at a c.m. energy, $E_{\rm cm}$, is determined with
\begin{equation}
\sigma^{\rm obs}(e^+e^-\rightarrow J/\psi{\rm X})
    =\frac{N^{\rm obs} - N_{\rm b} }
                {\mathcal L~{\epsilon}
                ~\mathcal B({J}/\psi \rightarrow \ell^+\ell^-)
                },
\label{eq:xsct_formula}
\end{equation}
where $N^{\rm obs}$ and
$N_{\rm b}$ are, respectively,
the number of  ${J}/\psi {\rm X}$ signal events 
obtained from the data
and the number of background events
estimated
by MC simulations,
$\mathcal L$ is the integrated luminosity of the data,
${\epsilon}$
is the efficiency for the
selection of $e^+e^-\rightarrow {J}/\psi{\rm X}$ events,
and $\mathcal B({J}/\psi \rightarrow \ell^+\ell^-)$
is the branching fraction for $J/\psi$ decays to the lepton pair $\ell^+\ell^-$.
To optimize the number of signal events, we do not fully reconstruct $X$.

The ${J}/\psi$ is reconstructed via the $e^+e^-$ and $\mu^+\mu^-$ final states. Each
event is required to have exactly two charged tracks and more than
one photon, or to have three or four charged tracks in the final state.
For each charged track, the polar angle $\theta$  in the
multilayer drift chamber (MDC)  must satisfy
$|\cos\theta|<0.93$.
For all charged tracks, the distance of closest approach to the average $e^+e^-$ interaction point 
is required to be less than 1.0~cm in the plane perpendicular
to the beam and less than 10.0~cm along the beam direction.
The electron and the muon can be well separated with the
ratio $E/p$, where $E$ is the energy deposited in the electromagnetic calorimeter (EMC)
and $p$ is the momentum of the charged track,
which is measured using the information in the MDC.
For $e^{\pm}$ candidates,  the ratio $E/p$ is required to be larger  than 0.7, while for $\mu^{\pm}$,  
it is required be be in the range from 0.05 to 0.35.  
To reject radiative Bhabha scattering events, the polar angles of
the leptons are required to satisfy $|\cos\theta|<0.81$,
and the angle between the two leptons to be less than 179$^{\circ}$.
The momenta of the leptons are required
to be larger than 1~GeV and less than $0.47\times E_{\rm cm}$.
To select $\pi^{\pm}$ 
and to reject backgrounds such as $\pi^+\pi^-K^+K^-$ from $c\bar c$ 
and non-$c\bar c$ state decays and two-photon exchange processes 
of $e^+e^-\rightarrow \ell^+\ell^-K^+K^-$,
the confidence level of the pion hypothesis,
calculated based on $dE/dx$ and time-of-flight measurements,
is required to be greater than that of the corresponding
kaon hypothesis.
For the selection of photons, the deposited energy of a neutral cluster in
the EMC is required to be greater than 25~MeV in barrel and
50~MeV in end-caps.
Time information from the EMC is used to
suppress electronic noise and energy deposits unrelated to the event.
To exclude fake photons originating from charged tracks,
the angle between the photon candidate and the nearest charged track
is required to be greater than $10^{\circ}$.

\begin{figure}[!hbt]
  \centering
\includegraphics[width=9.0cm,height=3.5cm]{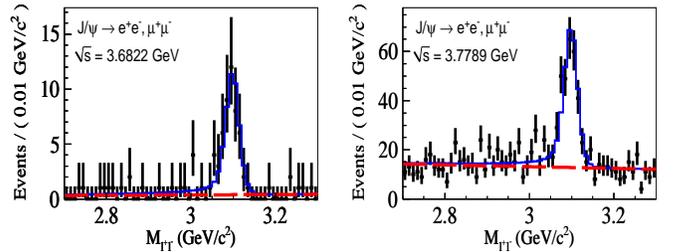}
\caption{
The invariant-mass distributions of the $\ell^+\ell^-$ pair selected from
data taken at two c.m. energies $E_{\rm cm}$,
where the dots with error bars are the number of the observed events,
the blue lines are the fit to these events,
while the dashed lines show the shape of the background.
  }
  \label{fig:MassOfLeptonPair_3773GeV}
\end{figure}

The numbers of candidates for $J/\psi$
are determined by fitting the $\ell^+\ell^-$ invariant mass spectra 
of the events satisfying the previously-described selection criteria.
This is illustrated in Fig.~\ref{fig:MassOfLeptonPair_3773GeV} which
shows the $\ell^+\ell^-$ invariant mass spectra from the data sets
obtained at two c.m. energies. Clear peaking structures can be
observed that stem from ${J}/\psi$ decays.
We fit these mass spectra with a function describing
both the signal and background shapes.
The signal shape is described by the MC-simulated signal shape,
while the smooth background is modeled by a line.
The fits yield the numbers
of the candidates for $e^+e^-\rightarrow {J}/\psi {\rm X}$.
Similarly, we obtain the numbers of the candidates for $e^+e^-\rightarrow {J}/\psi {\rm X}$
selected from all of the data sets taken at the other
energies.

These selected candidate events still contain some background events,
originating from several sources, which includes:
(1) $e^+e^-\rightarrow (\gamma)e^+e^-$,
(2) $e^+e^-\rightarrow (\gamma)\mu^+\mu^-$,
(3) $e^+e^-\rightarrow (\gamma)\tau^+\tau^-$,
(4) $e^+e^-\rightarrow (\gamma)D^+D^-$,
(5) $e^+e^-\rightarrow (\gamma)D^0 \bar D^0$,
(6) continuum light hadron production,
(7) $e^+e^-\rightarrow (\gamma){J}/\psi$ events.
Detailed MC studies of these backgrounds show that
only one major background source of $e^+e^-\rightarrow (\gamma){J}/\psi \rightarrow (\gamma)\ell^+\ell^-$
could be misidentified as $e^+e^-\rightarrow {J}/\psi {\rm X}$,
which is due to picking up fake photons or unphysical charged track(s).
From these MC studies we find that the fraction of these background events misidentified as signal events is
$\eta_{\rm mis}=(0.181 \pm 0.024)\%$.
With the ${J}/\psi$ resonance parameters~\cite{pdg2018} as inputs and
considering the energy spread, we extract the cross section
$\sigma^{\rm ISRV}_{e^+e^-\rightarrow (\gamma){J}/\psi}$ for $e^+e^-\rightarrow (\gamma){J}/\psi$, 
which include both the ISR and vacuum polarization effects, and we determine
$N_{\rm b} = {\mathcal L} \sigma^{\rm ISRV}_{e^+e^-\rightarrow (\gamma){J}/\psi} \eta_{\rm mis}$.

The efficiencies for the selection of
$e^+e^-\rightarrow {J}/\psi {\rm X}$
decays are determined with MC simulated events for these decays including
the ISR and final-state radiative effects,
where the final states include $J/\psi \pi^+\pi^-$, $J/\psi \pi^0\pi^0$, $J/\psi \eta$,
$J/\psi \pi^0$, $\gamma \chi_{cJ}$ ($J$=0,1,2) in which $\chi_{cJ}\rightarrow \gamma J/\psi$
followed by $J/\psi \rightarrow e^+e^-$ and $J/\psi \rightarrow \mu^+\mu^-$,
whose decay branching fractions are given by Particle Data Group~\cite{pdg2018}.
With the MC samples generated at 69
c.m. energies ranging
from 3.645 to 3.895~GeV, we determine the corresponding efficiencies.
We observe an energy-dependent efficiency curve increasing
smoothly from $58.8\%$ at 3.645~GeV to $60.8\%$ at 3.891~GeV. 
With the numbers of candidates for $e^+e^-\rightarrow {J}/\psi {\rm X}$
selected from the 69 data sets,
$N_b$, $\eta_{\rm mis}$, $\mathcal L$, ${\epsilon}_{{J}/\psi{\rm X}}$,
and $\mathcal B({J}/\psi \rightarrow \ell^+\ell^-)$,
we determine the observed cross sections at these energies, 
which are shown in Tab.~\ref{tab:ObsCS_At_Energies}
\begin{table*}
\centering
\caption{\sf Observed cross section of $e^+e^-\rightarrow J/\psi X$ at
energies from 3.6451 to 3.891 GeV.
Shown in the table are the center-of-mass energy $E_{\rm cm}$ and the observed cross section $\sigma^{\rm obs} ({\rm
nb})$, where the uncertainty is due to statistical fluctuation.}
\begin{tabular}{cccccc}
  \hline
  \hline
  $E_{\rm cm}(\rm GeV)$ & $\sigma^{\rm obs} ({\rm nb})$ & $E_{\rm cm}(\rm GeV)$ & $\sigma^{\rm obs} ({\rm nb})$
  & $E_{\rm cm}(\rm GeV)$ & $\sigma^{\rm obs} ({\rm nb})$ \\
  \hline
  3.6451    &    0.1584 $\pm$ 0.1467  &   3.7002    &    9.2087 $\pm$ 1.7830  &  3.7760    &    1.8234 $\pm$ 0.1509   \\
  3.6474    &    0.1839 $\pm$ 0.0791  &   3.7055    &    5.4001 $\pm$ 1.2753  &  3.7789    &    1.9568 $\pm$ 0.1490   \\
  3.6500    &    0.0353 $\pm$ 0.0170  &   3.7136    &    5.9442 $\pm$ 1.0105  &  3.7818    &    1.7823 $\pm$ 0.1374   \\
  3.6534    &    0.1011 $\pm$ 0.0756  &   3.7215    &    3.9790 $\pm$ 0.7401  &  3.7847    &    1.7938 $\pm$ 0.1254   \\
  3.6789    &    2.2886 $\pm$ 0.9609  &   3.7269    &    3.7609 $\pm$ 0.2893  &  3.7873    &    1.7704 $\pm$ 0.1201   \\
  3.6799    &    3.2183 $\pm$ 1.2272  &   3.7296    &    4.3698 $\pm$ 0.6063  &  3.7915    &    1.7365 $\pm$ 0.1062   \\
  3.6809    &    4.2526 $\pm$ 1.3193  &   3.7359    &    2.6123 $\pm$ 0.4197  &  3.7952    &    1.7198 $\pm$ 0.0987   \\
  3.6818    &    5.2955 $\pm$ 1.4177  &   3.7368    &    3.1582 $\pm$ 0.3642  &  3.7989    &    1.4748 $\pm$ 0.0976   \\
  3.6822    &   15.1674 $\pm$ 2.3084  &   3.7379    &    3.6116 $\pm$ 0.4794  &  3.8030    &    1.4094 $\pm$ 0.1161   \\
  3.6826    &   23.1799 $\pm$ 2.7139  &   3.7454    &    3.0186 $\pm$ 0.2544  &  3.8068    &    1.3579 $\pm$ 0.1445   \\
  3.6834    &   63.1031 $\pm$ 4.5335  &   3.7470    &    2.8673 $\pm$ 0.2058  &  3.8099    &    1.2778 $\pm$ 0.1617   \\
  3.6840    &  132.6346 $\pm$ 6.7087  &   3.7493    &    2.4755 $\pm$ 0.1529  &  3.8128    &    1.3236 $\pm$ 0.1961   \\
  3.6846    &  227.0219 $\pm$ 9.2357  &   3.7508    &    2.5454 $\pm$ 0.1358  &  3.8160    &    1.2392 $\pm$ 0.2304   \\
  3.6848    &  315.1179 $\pm$ 12.3896 &   3.7530    &    2.5439 $\pm$ 0.1294  &  3.8240    &    1.0511 $\pm$ 0.2713   \\
  3.6854    &  398.1874 $\pm$ 14.4367 &   3.7544    &    2.3480 $\pm$ 0.1231  &  3.8319    &    2.1713 $\pm$ 0.3873   \\
  3.6860    &  390.2015 $\pm$ 13.7130 &   3.7558    &    2.4779 $\pm$ 0.1187  &  3.8400    &    1.1356 $\pm$ 0.3131   \\
  3.6861    &  409.1987 $\pm$ 0.5692  &   3.7587    &    2.4333 $\pm$ 0.1086  &  3.8479    &    0.9437 $\pm$ 0.3123   \\
  3.6866    &  384.0587 $\pm$ 13.7238 &   3.7617    &    2.2402 $\pm$ 0.1053  &  3.8561    &    0.7033 $\pm$ 0.2532   \\
  3.6873    &  321.1585 $\pm$ 12.2803 &   3.7645    &    1.9580 $\pm$ 0.1164  &  3.8640    &    0.6309 $\pm$ 0.2489   \\
  3.6874    &  300.4553 $\pm$ 11.8697 &   3.7674    &    1.8666 $\pm$ 0.1316  &  3.8719    &    0.9599 $\pm$ 0.2243   \\
  3.6890    &  107.3780 $\pm$ 6.7288  &   3.7702    &    1.6418 $\pm$ 0.1398  &  3.8809    &    1.3120 $\pm$ 0.4021   \\
  3.6920    &   26.6893 $\pm$ 3.3631  &   3.7730    &    1.8966 $\pm$ 0.0041  &  3.8909    &    0.7580 $\pm$ 0.3471   \\
  3.6964    &   13.6963 $\pm$ 2.2532  &   3.7731    &    1.6579 $\pm$ 0.1483  &  3.8077    &    1.3816 $\pm$ 0.0250   \\
  \hline
  \hline
\end{tabular}
\label{tab:ObsCS_At_Energies}
\end{table*}

For estimating the systematic uncertainty of the cross-section measurements, we considered 12 sources,
most of which are determined by comparing the corresponding quantities obtained from both data and MC
simulated events.
In the following, we summarize the various contributions and our estimate 
of the respective systematic error in parenthesis:
(1) the uncertainty in the efficiency determination due to the angle 
$\theta_{\ell^+\ell^-}$ cut for the leptons (negligible),
(2) due to the $\cos\theta$ cut for the charged tracks ($0.4\%$),
(3) due to the $E/p$ cut ($0.3\%$),
(4) due to the lepton momentum $p_{l^{\pm}}$ cut ($0.2\%$),
(5) due to the constraints applied on the number of charged tracks cut or photons ($0.4\%$),
(6) the uncertainty of the tracking efficiency ($0.3\%$) for pions, 
    while this uncertainty for the leptons cancels with the corresponding uncertainty 
    in the luminosity measurements,
(7) the uncertainty induced by fitting the invariant-mass spectrum ($0.8\%$),
(8) the uncertainty in the modeling of the MC 
    ($0.9\%$) including the
    branching-fraction uncertainties of
     $\psi(3686)\rightarrow {J}/\psi \pi^+\pi^-$ and $\psi(3686)\rightarrow {J}/\psi \pi^0\pi^0$;
(9) the uncertainty related to the identification of $\pi^{\pm}$ ($1.0\%$),
(10) the uncertainty in the branching fraction for the decay ${J}/\psi \rightarrow l^+l^-$
    ($0.4\%$)~\cite{pdg2018},
(11) the uncertainty in the background subtraction for the decay
    $e^+e^-\rightarrow {J}/\psi {\rm X}$ ($<0.1\%$)
    due to uncertainty of ${J}/\psi$ resonance parameters, and
(12) the systematic uncertainty in the luminosity measurements ($1.0\%$).
    Adding the individual systematic uncertainties in quadrature, assuming them to be independent,
    yields a total systematic uncertainty of $2.0\%$.

\begin{figure}[!hbt]
  \centering
\vspace{-0.0cm}
\includegraphics[width=8.0cm,height=5.5cm]{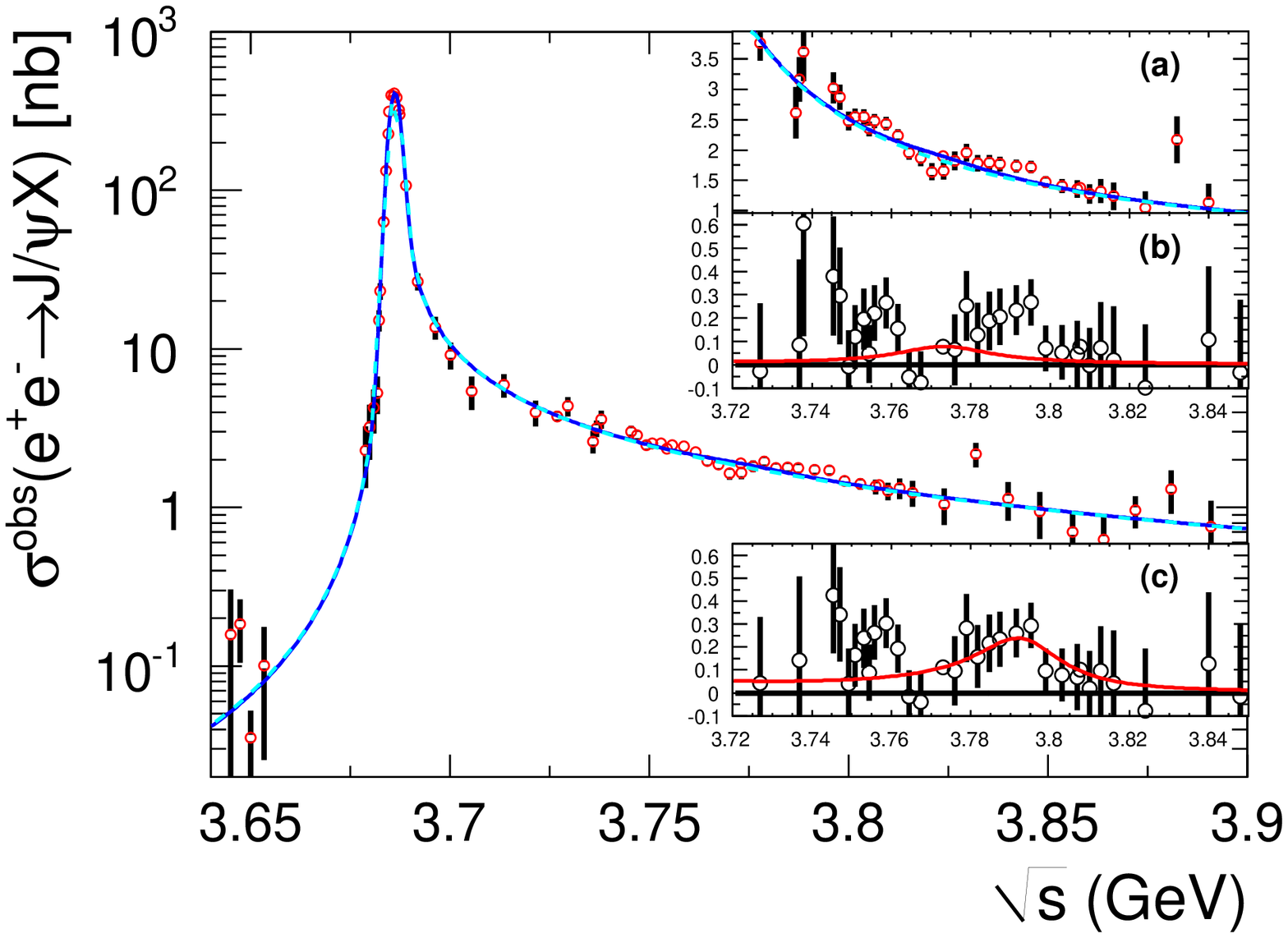}
  \caption{
The observed cross section for $e^+e^-\rightarrow {{J}/\psi{\rm X}}$
and the best fit to the cross section under the assumption
that the $\psi(3686)$ and $\psi(3770)$ decays contribute to the cross
section.
}
  \label{fig:Fit_to_Xscts_With_P3686P3770}
\end{figure}

Figure~\ref{fig:Fit_to_Xscts_With_P3686P3770} shows the observed cross sections as circles with error bars,
where the errors are statistical taking into account statistical fluctuations of the signal, the number of MC events,
and the statistical uncertainties of the luminosity measurements.
The dominant peak located at $\sim$3.686~GeV is due to $\psi(3686)$ decays.
The shape of the cross section at energies above 3.73~GeV is
anomalous, indicating that there could be structure(s)
lying at energies 
from 3.73 to 3.87~GeV
as those observed by the BES Collaboration~\cite{bes2_prl_2structures}.

We analyze the cross section by performing least-$\chi^2$ fits to the cross section.
The expected cross section can be modeled with
\begin{equation}
\begin{tiny}
 \sigma^{\rm exp}(s) =
  \int_{\sqrt{s}_{-}}^{\sqrt{s}_{+} }
   dw {\mathcal G(s,w)}\int_{0}^{1-\frac{M^2_{J/\psi} }{s} }
   dx {\sigma}^{\rm dress}(s')
{\mathcal F(x,s)},
\end{tiny}
\end{equation}
\noindent
where
$s=E^2_{\rm cm}$, $x$ is the energy fraction of the radiative photon~\cite{Structure_Function},
$s'=s(1-x)$, ${\mathcal G(s,w)}$~\cite{PhysRevLetts97_121801_2006} represents a Gaussian
function~\cite{PhsLettB652_238_2007}
describing the c.m. energy distribution of BEPCII,
$w$ is an integration variable,  
$\sqrt{s}_{\pm}=\sqrt{s}\pm 5\Delta_{\rm sprd}$, in which
$\Delta_{\rm sprd}$ is the energy spread, $\sigma^{\rm dress}(s')$
is the dressed cross section 
including vacuum polarization effects 
for the ${J}/\psi{\rm X}$ production, $M_{J/\psi}$ is the mass of $J/\psi$,
and ${\mathcal F(x,s)}$ is a sampling function~\cite{Structure_Function}.

We perform the least-$\chi^2$ fits
with two hypotheses of
$\sigma^{\rm dress}(s') = |A_{\psi(3686)}(s') + e^{i\phi_1}A_{\mathcal S}(s')|^2$,
and
$\sigma^{\rm dress}(s') = |A_{\psi(3686)}(s') + e^{i\phi_1}A_{R_1}(s') +e^{i\phi_2}A_{R_2}(s')|^2$,
separately, 
where $A_{\psi(3686)}(s')$, $A_{\mathcal S}(s')$, $A_{R_1}(s')$ and
$A_{R_2}(s')$ are, respectively,
the decay amplitudes of $\psi(3686)$,
${\mathcal S}$ (${\mathcal S}$ could be either $\psi(3770)$ or a new structure),
$R_1$, and $R_2$, while $\phi_1$ and $\phi_2$ are the corresponding phases of these amplitudes.
The generic decay amplitude of the resonance or structure are described by a
Breit-Wigner function
$A_j(s') = { \sqrt{12 \pi \Gamma^{ee}_j \Gamma^{\rm tot}_j {\mathcal B_j}}}
          /{[(s'-M_j^2) +iM_j\Gamma^{\rm tot}_j]}$,
where the subscript $j$ indicates one of these resonances, $M_j$, $\Gamma^{ee}_j$, $\Gamma^{\rm tot}_j$,
and ${\mathcal B_j}$ represent the mass, leptonic width, total width,
and branching fraction of the $j$ resonance or structure decaying into  ${J}/\psi{\rm X}$ final
states, respectively.

In the fit the observed cross-section values
are assumed to be influenced only by the
uncertainties of statistical origin~\cite{BESIII_eeToMuMu_arXiv2007p12872v1_hep_ex_25Jul2020}.
The uncertainties on
the parameters returned by the fit 
are referred to as statistical uncertainties
in the subsequent discussion.
The remaining cross-section
uncertainties (assumed to be fully correlated
between different energies) are taken into account
using the ``offset method''~\cite{offset}.
The cross-section values are changed 
for all energies simultaneously by the 
size of the uncertainty and the resulting change
in the fit parameter is taken
as a systematic uncertainty.

The solid line in Fig.~\ref{fig:Fit_to_Xscts_With_P3686P3770}
shows the best fit result  under assumption of the two decay amplitudes
contributing to the cross sections,
while the dashed line
shows the contribution
from $\psi(3686)\rightarrow {J}/\psi {\rm X}$ decays.
To highlight the difference between the measured cross sections and their fitted
values, three enlarged figures are inserted.
The sub-figure (a)
shows the cross section with the fit,
while the sub-figure (b) shows the cross section
with the fit, where the $\psi(3686)$
contributions are subtracted. The solid line in the sub-figure (b)
corresponds to the best fit result of the cross section taking into account the $\psi(3770)$ decay and
interference effects between the $\psi(3686)$ and $\psi(3770)$ decay
amplitudes.
In this fit, the total and leptonic widths of both the $\psi(3686)$ resonance and
${\mathcal S}$
structure as well as the mass of
${\mathcal S}$
are fixed to the values of these for $\psi(3686)$ and $\psi(3770)$
given by the Particle Data Group~\cite{pdg2018},
while the branching fractions for the decays of ${\psi(3686)\rightarrow {J}/\psi{\rm X}}$
and ${\psi(3770)\rightarrow {J}/\psi{\rm X}}$
as well as the phase $\phi_1$ are left as free parameters.
The fit has two solutions for one free phase.
The two solutions have an identical fit goodness
of $\chi^2=120.4$ 
for 64 degrees of freedom.
Table~\ref{tab:Comp_FitResults_Psi3686S3760BESII} shows results from the fits,
where the first uncertainties are statistical, and second systematic.
The systematic uncertainties have three sources for
${\mathcal B}(\psi(3686)\rightarrow {J}/\psi{\rm X})$,
${\mathcal B}(\psi(3770)\rightarrow {J}/\psi{\rm X})$ and
$\phi_1$, which are respectively:
 (1) $2.33\%$, $1.96\%$ and $0.01\%$ for them
due to uncertainty of the observed cross section,
(2) $0.16\%$, $9.80\%$ and $6.01\%$
due to uncertainties of the fixed parameters, and
(3) $0.62\%$, $27.45\%$ and $0.86\%$
due to uncertainties of $E_{\rm cm}$. Adding these uncertainties in quadrature
yields the total systematic uncertainty. The branching fraction for the decay
$\psi(3770)\rightarrow {J}/\psi{\rm X}$ from one solution is consistent within error
with the sum of published branching fractions~\cite{pdg2018},
$(0.47 \pm 0.06)\%$~\cite{pdg2018}, of $\psi(3770)\rightarrow J/\psi \pi^+\pi^-$, $J/\psi \pi^0\pi^0$,
and $J/\psi \eta$, $\gamma \chi_J$ with $J=0,1,2$,
while the branching fraction from the other solution is larger than the
total branching fraction by a factor of about four.

If we leave the mass and total width of
${\mathcal S}$
free in the fit, the fit returns
$\chi^2$$=$$95.4$ 
for 62 degrees of freedom, 
with a mass, total width and product of branching fraction and
leptonic width of $3793.5$$\pm2.0$$\pm1.6$ MeV$/c^2$, $25.3$$\pm9.1$$\pm2.4$ MeV and
$8.0$$\pm3.2$$\pm2.0$ eV respectively,
where the first uncertainties are statistical, and second systematic (see below).
The inserted sub-figure (c) in Fig. 2 shows an enlarged plot of the cross sections
in which the $\psi(3686)$ contributions are subtracted.
For convenience, we call the structure 
${\mathcal S(3790)}$.
The mass 
$M_{\mathcal S(3790)}=$$3793.5$$\pm2.0$$\pm1.6$ MeV$/c^2$
deviates
$7.6$ times of standard deviation from
$M_{\psi(3770)}$=$3773.13$$\pm0.35$ MeV$/c^2$~\cite{pdg2018},
indicating that 
${\mathcal S(3790)}$
observed in $J/\psi{\rm X}$ final state is not
$\psi(3770)$.
Removing the ${\mathcal S(3790)}$ 
from the fit yields the fit
$\chi^2=139.1$. Reducing four degrees of freedom in the fit causes the fit
$\chi^2$ change by 43.7, indicating that the signal significance of 
${\mathcal S(3790)}$
is $5.8\sigma$.
\begin{center}
{\vspace{-0.6cm}}
\begin{table}[!hbp]
\centering
\caption{
Results of fits as described in the text,
where ${\mathcal B_1}$ and ${\mathcal B_2}$ indicate, respectively, the branching fractions
of $\psi(3686)$$\rightarrow$${J}/\psi{\rm X}$ and
$\psi(3770)$$\rightarrow$${J}/\psi{\rm X}$ decays, and $\phi_{\mathcal R_1}$
is the phase of the amplitude $A_{\mathcal S}(s')$.
}
\small
\begin{tabular}{ccc}
\hline\hline
Solution & Solution I  & Solution II  \\
\hline
${\mathcal B_1}$                 & $(64.4\pm0.6\pm1.4)\%$    & $(64.6\pm 0.6\pm 1.4)\%$ \\
${\mathcal B_2}$                & $(0.51\pm0.17\pm 0.15)\%$ & $(2.21\pm0.46\pm 0.64)\%$  \\
$\phi_{\mathcal R_1}$ & $(93.2\pm 52.2\pm 5.7)^{\circ}$ & $(-105.0\pm 24.8\pm 6.2)^{\circ}$ \\
\hline
\hline
\end{tabular}
\label{tab:Comp_FitResults_Psi3686S3760BESII}
\end{table}
\end{center}
\begin{figure}[!hbt]
  \centering
\vspace{-0.4cm}
\includegraphics[width=8.0cm,height=5.5cm]{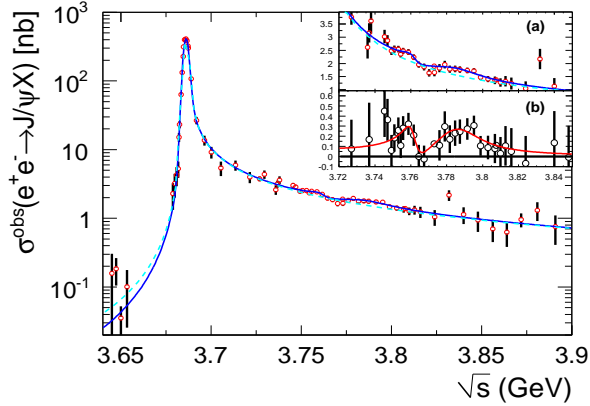}
\caption{
 The observed cross section for $e^+e^-\rightarrow {{J}/\psi{\rm X}}$
and the best fit to the cross section under the assumption
that the $\psi(3686)$, $R(3760)$ and $R(3790)$ decays contribute to the cross
section.
  }
  \label{fig:Fit_to_Xscts_With_P3686S3760S3780}
\end{figure}

In the inserted sub-figure (c) in Fig. 2, the observed cross section values around
3.758 GeV significantly deviate from the fitted values of the cross
section, which could be due to a new structure produced around this energy. 
To interstage whether these are due to
the new structure, we add one more BW amplitude in the fit leaving four
parameters free.
Figure~\ref{fig:Fit_to_Xscts_With_P3686S3760S3780}
shows the fit to the cross-section data
under the assumption of the presence of the three contributing decay amplitudes,
where the solid line shows the best fit, the dashed line
shows the contribution from $\psi(3686)\rightarrow {J}/\psi {\rm X}$ decays,
the inserted sub-figure (a) shows the cross section
with the best fit in the energy range from 3.72 to 3.85 GeV,
and the sub-figure (b) shows the cross section with the best fit, in which
the $\psi(3686)$ contributions are subtracted.
Table~\ref{tab:FitResults_Psi3686S37xxS37yy}
summarizes the results from the fits, where the first uncertainties
are statistical, and the second systematic. 
The values in brackets are the upper limits of the values 
of the parameters set at $90\%$ C.L..

The systematic uncertainties in the magnitude of the fitted parameters
are assumed to originate from four sources:
(1) the uncertainty of the observed cross section,
(2) the uncertainties of the total and leptonic widths of $\psi(3686)$,
(3) the uncertainties of the c.m. energies,
and
(4) the uncertainties of the branching fractions of $R(3760)$ and $R(3790)$ decays.
To estimate these uncertainties, we change the values of the fixed parameters by $\pm 1\sigma$, and
refit the observed cross section, and subsequently take the difference between the refitted parameters
and the ones of the nominal fit result as the corresponding systematic shift.
A similar procedure has been applied to estimate the systematic error related
to uncertainties of the c.m. energies. In this case, we vary the c.m. energies
with a Gaussian uncertainty of 0.25~MeV in the resonance energy region,
refit the data, and take the difference of the updated fit parameters with respect
to the result of the nominal fit as a measure of the systematic shift.
Taking these shifts as independent systematic uncertainties, we add them in quadrature
to obtain the total systematic uncertainty for each parameter, which is indicated by the
second uncertainty of each parameter in Table~\ref{tab:FitResults_Psi3686S37xxS37yy}.
\begin{table}[!hbp]
\caption{
The fitted results, where $M$, $\Gamma^{\rm tot}$ and $\Gamma^{ee}$ are the
mass, total, and leptonic widths of resonance(s) $\mathcal R_i$,
$\mathcal B({R_i}\rightarrow f)$ ($f=J/\psi{\rm X}$) is the branching fraction for the $\mathcal R_i$
decay into $f$, and $\phi_i$ is the phase of the amplitude,
in which $i=1,2,3$ indicate $\psi(3686)$, $R(3760)$ and $R(3790)$, respectively;
$\mathcal B(R_2\rightarrow f)$
and $\mathcal B(R_3\rightarrow f)$
are determined using $\Gamma^{ee}_{R(3760)}=186\pm 201\pm 8$~eV and
$\Gamma^{ee}_{R(3790)}=243\pm 160\pm 9$~eV~\cite{bes2_prl_2structures},
where the normalization uncertainties are not included.
}
\begin{tabular}{lcr}
\hline\hline
Parameter            &  Solution I                & Solution II \\
\hline
$\mathcal B(\mathcal R_1\rightarrow f)$ [$\%$]
        & $62.7\pm 0.2\pm 1.3$   & $62.4\pm 0.1\pm 1.3$   \\
${M}_{\mathcal R_2}$ [MeV/$c^2$]
        & $3761.7\pm 2.2\pm 1.2$ & $3762.0\pm 1.7\pm 1.2$ \\
$\Gamma^{\rm tot}_{\mathcal R_2}$ [MeV]
        & $ 6.7\pm 11.1\pm 1.1$  & $7.1\pm 4.4\pm 1.1$  \\
        & ($<21.0$)  &  ($<12.9$)    \\
$\Gamma^{ee}_{\mathcal R_2}\mathcal B(R_2\rightarrow f)$ [eV]
        & $4.0\pm 4.3\pm 1.2$    & $3.8\pm 3.7\pm 1.1$ \\
        & ($<9.7$)     &  ($<8.7$)    \\
$\phi_1$ [degree]
        & $279.2\pm 53.7\pm 9.0$ & $208.1\pm 34.5\pm 5.2$
\\
${M}_{\mathcal R_3}$ [MeV/$c^2$]
        & $3784.7\pm 5.7\pm 1.6$ & $3784.3\pm 4.9\pm 1.5$
\\
$\Gamma^{\rm tot}_{\mathcal R_3}$ [MeV]
        & $ 31.6\pm 11.9\pm 3.2$ & $32.7\pm 8.9\pm 3.2$ \\
        & ($<47.4$)              &                      \\
$\Gamma^{ee}\mathcal B({\mathcal R_3\rightarrow f})$ [eV]
        & $18.1\pm 10.3\pm 4.7$  & $12.6\pm  4.6\pm  3.2$  \\
        & ($<33.9$)              &  ($<19.8$)    \\   
$\phi_2$ [degree]
        & $209.6\pm27.3\pm 19.9$ &  $115.5\pm40.0\pm 10.9$
\\
$\mathcal B(\mathcal R_2\rightarrow f)$ [$\%$]
      & $2.1\pm 2.3\pm 0.6$ & $2.0\pm 2.0\pm 0.6$ \\
      & ($<5.1$)              &  ($<4.7$)    \\
$\mathcal B({\mathcal R_3\rightarrow f})$ [$\%$]
      & $7.4\pm 4.3\pm 1.9$ & $5.2\pm 1.9\pm 1.2$ \\
      & ($<13.4$)              &  ($<8.1$)    \\
\hline
\hline
\end{tabular}
\label{tab:FitResults_Psi3686S37xxS37yy}
\end{table}

The fit has four solutions for two free phases.
However, we only found two distinguishable solutions.
Two of the solutions overlap with the other two, as expected according to
mathematical predictions reported in Ref.~\cite{PRD99_072007_Y2019_Int_J_Mod_Physics_A26_4511_Y2011}.
The two distinct solutions, summarized in
Table~\ref{tab:FitResults_Psi3686S37xxS37yy},
gave a fit quality of $\chi^2$=$78.6$ with 58 degrees of freedom.
Thus, both solutions are equally acceptable.
We choose Solution I as the nominal results of the analysis.
The fit yields a branching fraction
$\mathcal B(\psi(3686)\rightarrow {J}/\psi{\rm X})=(62.7\pm 0.2 \pm 1.5)\%$,
where the first uncertainty is statistical, and the second systematic.
This result is consistent within error with the world average of
$\mathcal B(\psi(3686)\rightarrow {J}/\psi{\rm X})$ $=(61.6\pm 0.6)\%$~\cite{pdg2018}.

Comparing the hypothesis of 
$\psi(3686)$+${\mathcal S(3790)}$
with all parameters of 
${\mathcal S(3790)}$
structure free in the fit, including an additional 
structure $R(3760)$ with all parameters free, 
reduce the $\chi^2$ of the fit by 16.8, which corresponds to a statistical
significance for the observation of the $R(3760)$ of $3.1\sigma$.
The masses and total widths 
of these two structures, $R(3760)$ and $R(3790)$,  
which are measured in this work are consistent 
within $1.3\sigma$ uncertainties with those
measured in analysis of $e^+e^- \rightarrow {\rm hadrons}$ by the BES
Collaboration~\cite{bes2_prl_2structures}.

    In summary, we have measured for the first time the observed cross sections 
for $e^+ e^- \to J/\psi X$ at c.m. energies ranging from 3.645 to 3.891 GeV. 
We fitted to the cross sections with the sum of the known $\psi(3686$) and $\psi(3770)$ 
states and obtained the branching fractions for their inclusive decays to $J/\psi X$
for the first time. 
The fitting quality can be improved by replacing $\psi(3770)$ with a float 
resonance $S(3790)$, 
but the yielded mass $M_{\mathcal S(3790)}=$$3793.5$$\pm2.0$$\pm1.6$ MeV$/c^2$
deviates $7.6$ times of standard deviation from 
$M_{\psi(3770)}=3773.13\pm0.35$ MeV$/c^2$~\cite{pdg2018}, 
indicating that ${\mathcal S(3790)}$ observed in $J/\psi{\rm X}$ 
final state is not $\psi(3770)$.
The statistical significance of this $S(3790)$ state is estimated to be 5.8$\sigma$ 
over the hypothesis of only $\psi(3686)\rightarrow J/\psi X$ contributing of the cross sections.  
In addition, we fitted the cross sections with two float resonances 
to replace $\psi(3770)$, whose statistical significance is evaluated 
to be 3.1$\sigma$ over the 
one float state hypothesis. 
Using the leptonic widths of the two structures measured by the BES
Collaboration~\cite{bes2_prl_2structures} at the BESII experiment as inputs,
we have determined the decay branching fractions of
$\mathcal B[R(3760)\rightarrow {J}/\psi{\rm X}]=(2.1\pm 2.3\pm 0.6\pm 2.4)\%$, and
$\mathcal B[R(3790)\rightarrow {J}/\psi{\rm X}]=(7.4\pm 4.3\pm 1.9\pm 4.9)\%$,
where the first uncertainties are statistical, the second ones systematic, and
the third ones are due to uncertainties of
the leptonic widths~\cite{bes2_prl_2structures}.
The fitted resonance parameters of the two structures are consistent with those 
of the di-structure~\cite{bes2_prl_2structures}.

\vspace{-1mm}

The BESIII collaboration thank the staff of BEPCII and the IHEP computing
center for their strong support. This work is supported in part by National
Key Basic Research Program of China under Contract No. 2009CB825204, 2015CB856700;
National Natural Science Foundation of China (NSFC) under Contracts Nos.
10935007,
11625523, 11635010, 11735014, 11822506, 11835012, 11961141012; the Chinese
Academy of Sciences (CAS) Large-Scale Scientific Facility Program; Joint
Large-Scale Scientific Facility Funds of the NSFC and CAS under Contracts
Nos. U1532257, U1532258, U1732263, U1832207; CAS Key Research Program of
Frontier Sciences under Contracts Nos. QYZDJ-SSW-SLH003, QYZDJ-SSW-SLH040;
100 Talents Program of CAS; 
CAS Other Research Program under Code No. Y129360;
INPAC and Shanghai Key Laboratory for Particle
Physics and Cosmology; ERC under Contract No. 758462; German Research
Foundation DFG under Contracts Nos. Collaborative Research Center CRC 1044,
FOR 2359; Istituto Nazionale di Fisica Nucleare, Italy; Ministry of
Development of Turkey under Contract No. DPT2006K-120470; National Science
and Technology fund; STFC (United Kingdom); The Knut and Alice Wallenberg
Foundation (Sweden) under Contract No. 2016.0157; The Royal Society, UK
under Contracts Nos. DH140054, DH160214; The Swedish Research Council; U. S.
Department of Energy under Contracts Nos. DE-FG02-05ER41374, DE-SC-0010118,
DE-SC-0012069.

\end{document}